# Bulk High-Entropy Hexaborides

Mingde Qin [a], Qizhang Yan [a], Yi Liu [a], Haoren Wang [a], Chunyang Wang [b], Tianjiao Lei [c], Kenneth S. Vecchio [a], Huolin L. Xin [b], Timothy J. Rupert [c], Jian Luo [a, *]

[a] *Department of NanoEngineering; Program of Materials Science and Engineering, University of California, San Diego, La Jolla, CA, 92093, USA*
[b] *Department of Physics and Astronomy, University of California, Irvine, Irvine, CA, 92697, USA*
[c] *Department of Materials Science and Engineering; Department of Mechanical and Aerospace Engineering, University of California, Irvine, Irvine, CA, 92697, USA*

**Abstract:**

For the first time, a group of $CaB_6$-typed cubic rare earth high-entropy hexaborides have been successfully fabricated into dense bulk pellets (>98.5% in relative densities). The specimens are prepared from elemental precursors via in-situ metal-boron reactive spark plasma sintering. The sintered bulk pellets are determined to be single-phase without any detectable oxides or other secondary phases. The homogenous elemental distributions have been confirmed at both microscale and nanoscale. The Vickers microhardness are measured to be 16-18 GPa at a standard indentation load of 9.8 N. The nanoindentation hardness and Young's moduli have been measured to be 19-22 GPa and 190-250 GPa, respectively, by nanoindentation test using a maximum load of 500 mN. The material work functions are determined to be 3.7-4.0 eV by ultraviolet photoelectron spectroscopy characterizations, which are significantly higher than that of $LaB_6$.

---

[*] Corresponding author. E-mail address: jluo@alum.mit.edu (J. Luo).

**Highlights:**

- High-entropy hexaborides were fabricated in the bulk form for the first time.
- A novel boron-metal reactive SPS processing resulted in >98.5 relative densities.
- The novel processing produced high-entropy hexaborides free of oxide inclusion.
- Elemental distributions are homogenous at both microscale and nanoscale.
- Microhardness, nanohardness, modulus, and work function were measured.

## 1. Introduction

In 2004, Yeh et al. [1] and Cantor et al. [2] published two independent reports on the fabrication of high-entropy alloys (HEAs) or multi-principal element alloys. In the last decade, HEAs have attracted great research interest in the metallurgy community. HEAs usually refer to metallic alloys with five or more elements of equimolar (or near equimolar) compositions, which can be considered as a subclass of complex concentrated (or compositionally complex) alloys (CCAs) [3]. HEAs and CCAs can possess superior, and sometimes unexpected, mechanical and other properties [3-6].

More recently, high-entropy ceramics (HECs), the ceramic counterparts to the metallic HEAs, have been fabricated in the bulk form, including rocksalt [7], perovskite [8], fluorite [9], pyrochlore [10], and spinel [11] oxides, borides [12-16], silicides [17, 18], carbides [19, 20], nitrides [21, 22], sulfides [23], and fluorides [24], as well as aluminides (intermetallics) [25]. The fabrication and properties of HECs have been summarized in several recent reviews [26-28]. Similar to HEAs and CCAs, it was recently proposed to generalize HECs to compositionally complex ceramics (CCCs), where medium-entropy or non-equimolar compositions sometimes possess better properties [10, 27, 29, 30].

Among the HECs, high-entropy borides (HEBs) offer abundant design opportunities due to their enormous structural, in addition to the high-entropy compositional, spaces. Based on a 2017 review by Akopov et al. [31], there are as many as 1253 entries (different crystal structures) for binary boron compounds of various stoichiometry ratios ranging from 4:1 to 1:66 ($M_4B$ to $MB_{66}$, where M represents a metal). In 2016, Gild et al. [12] reported the first class of HEBs, high-entropy metal diborides in the $AlB_2$-typed hexagonal structure. Further studies fabricated monoborides [14, 32], $M_3B_4$ borides [15], and tetraborides [16] in bulk form by using boro/carbothermal reduction of metal oxides [13, 33] or reactive sintering of elemental precursors [14-16, 32].

$CaB_6$-typed cubic hexaborides (Space group: Pm-3m, No. 221), including $YB_6$, $LaB_6$, $CeB_6$, $PrB_6$, $NdB_6$, $SmB_6$, $EuB_6$, $GdB_6$, $TbB_6$, $DyB_6$, $HoB_6$, $ErB_6$ and $YbB_6$, have been reported to be stable [34, 35]. The hexaboride structure is schematically illustrated in Fig. 1 (that it is drawn for a high-entropy hexaboride with five metals). In its cubic unit cell, metal cations locate at the body center, while boron octahedra (composed of six boron atoms) situate at eight vertices. A recent review [36] summarized four synthesis routes for rare earth hexaborides: (1) solid-state reactions, (2) melt electrolysis and flux, (3) vapor deposition and metal-gas reactions, and (4) combustion synthesis, while hot-press and spark plasma sintering (SPS) are the most popular densification methods. In general, rare earth hexaborides show great potential for a range of applications, including electron emitters, thermoelectric materials, coatings, and superconductors [35, 36]. Specifically, $LaB_6$ is extensively studied and widely used as a thermionic electron emitter due to its low work function and low vapor pressure at high temperatures [37]. The close lattice parameters ranging from 4.069 Å ($HoB_6$) to 4.176 Å ($EuB_6$) [34, 35] suggested they are easy to form solid solutions, where their properties can be further tuned.

High-entropy hexaborides (HEHBs; Fig. 1) have been successfully synthesized, but only as powders [33] or highly porous materials [13] (but not yet in the dense bulk form). Recently, Qin et al. showed a generic in-situ metal-boron reactive SPS method can be used to synthesize and fabricate (in one process) dense bulk pellets of four different classes of HEBs (in the MB, $MB_2$, $M_3B_4$, and $MB_4$ stoichiometries) [14-16, 38]. This motivates us to adopt this direct metal-boron reactive SPS route to fabricate HEHBs in bulk forms and subsequently measure their basic bulk properties.

## 2. Experimental Procedure

Commercial powders of Y, La, Pr, Nd, Sm, Gd, Tb, Dy, and Yb (99.9% purity, ~40 mesh, purchased from Alfa Aesar, MA, USA) and boron (99% purity, 1-2 μm, purchased from US Research Nanomaterials, TX, USA) were utilized as elemental precursors for synthesizing bulk specimens of six compositions HEHB1 to HEHB6 listed in Table 1. For each composition, elemental powders were weighted out following the nominal stoichiometric ratios in batches of 5 g. Then, the powders were first mixed by a vortex mixer, and successively high energy ball milled (HEBM) in a SPEX 8000D mill (SPEX CertiPrep, NJ, USA). Tungsten carbide lined stainless steel jars and 11.2 mm tungsten carbide milling media (at a ball-to-powder ratio of 4.5:1) were utilized during the HEBM process of 50 min, and 1 wt. % (0.05 g) of stearic acid was added as lubricant. After HEBM, the powder mixtures were loaded into 10 mm SPS graphite dies lined with graphite foils in batches of 2.5 g, and consecutively sintered into bulk pellets with a Thermal Technologies 3000 series SPS (Thermal Technology LLC, CA, USA) in vacuum ($10^{-2}$ Torr). To prevent oxidation, the HEBM and loading/handing of as-milled powder mixtures were both conducted in an argon atmosphere with $O_2$ < 10 ppm. The SPS sintering procedure is similar to that used in our previous work for fabricating rare earth high-entropy tetraborides (HETBs) [16] with isothermal holding at 1700 °C and 50 MPa for 10 min during the final densification.

After cooling down in the SPS machine, the sintered specimen pellets were ground (to remove surface carbon contamination) and successively mirror polished before characterizations. X-ray diffraction (XRD) characterizations were performed at 30 kV and 15 mA with Cu Kα radiation on a Rigaku Miniflex diffractometer. Specimen densities were directly measured via the Archimedes method. The theoretical densities were calculated from the nominal compositions and the lattice parameters measured by XRD. Scanning electron microscopy (SEM), energy dispersive X-ray spectroscopy (EDS), and electron backscatter diffraction (EBSD) were all conducted using a Thermo-Fisher (FEI) Apreo microscope equipped with an Oxford N-Max$^N$ EDS detector and an Oxford Symmetry EBSD detector. Transmission electron microscopy (TEM) samples of both HEHB1 and HEHB3 were prepared by focused-ion beam (FIB) milling on Thermo-Fisher (FEI) dual-beam FIB/SEM and ion milling on Fischione Model 1040 NanoMill. Atomic-resolution annular dark-filed scanning transmission electron microscopy (ADF-STEM) and nanoscale EDS were performed on a JEOL 300CF microscope and Talos F200X.

Vickers microhardness tests were carried out on a LECO hardness testing machine equipped with a diamond microindentor at a standard loading force of 9.8 N (1 kgf) with a holding time of 15 s, following the ASTM Standard C1327. More than 30 indention tests were performed at different locations on each specimen to minimize the microstructural and grain boundary effects and ensure statistical validity. The nanoindentation tests were performed on polished surfaces with an Agilent Nano Indenter G200. A diamond Berkovich indenter was used, and the maximum load was set to 500 mN. For each specimen, a six-by-six array with a distance of 30 μm between the two nearest indents was employed.

Ultraviolet photoelectron spectroscopy (UPS) measurements were conducted on freshly polished surface of each specimen (to avoid any surface oxidation) using a Kratos AXIS Supra spectrometer equipped with a He I ($hv$ = 21.22 eV) source under $10^{-8}$ Torr chamber pressure. The work functions were subsequently calculated based on the incident photon energy ($hv$) and the high-binding-energy cut-offs ($E_{\text{cut-off}}$) of each specimen via $WF = hv - E_{\text{cut-off}}$.

## 3. Results and Discussion

Fig. 2 illustrates the XRD patterns of all six compositions (listed in Table 1) fabricated via in-situ metal-boron reactive SPS. All six synthesized specimens demonstrate a single $CaB_6$-typed cubic hexaboride phase without any detectable oxides or any other secondary phase. Based on unit cell refinement of the measured XRD patterns, HEHB lattice parameters are calculated and listed in Table 1. Averaged lattice parameters calculated via the rule of mixture (RoM) of individual binary hexaborides (listed in Supplementary Table S1) are also tabulated in Table 1. In all six cases, there are negligible differences (~0.1%) between the XRD measured lattice parameters and RoM averages. The calculated XRD patterns based on the nominal compositions and measured lattice parameters are also displayed in Supplementary Fig. S1 for comparison. By the combination of XRD patterns and lattice parameters, it can be clearly observed that single-phase $CaB_6$-typed structures have been successfully obtained in all six compositions studied.

The formation of single high-entropy phase can be understood because of (1) high thermal stability of binary hexaborides [39, 40] and (2) extended homogeneity range at boron-rich end of hexaboride structures [40]. For the binary hexaborides involved in this study, they will either melt congruently ($LaB_6$, $PrB_6$, $NdB_6$, $SmB_6$, and $YbB_6$) or dissociate into tetraboride and boron-rich liquid phase ($YB_6$, $GdB_6$, $TbB_6$, and $DyB_6$) at high temperatures [34]. Except for $TbB_6$, $DyB_6$, and $YbB_6$ that melt/dissociate within a temperature range of 2200~2400 °C, all constitutive binary hexaborides can maintain thermodynamic stability until the temperature is higher than 2500 °C based on the phase diagrams in Ref. [40]. At the same time, hexaborides are known to possess extended homogeneity range at the boron-rich boundary by formation of metal vacancies [40]. Studies have demonstrated the stability of the hexaboride structure (mainly the boron sublattice) to a metal content as low as $Sm_{0.68}B_6$ and $Pr_{0.69}B_6$ [41, 42]. On the other hand, the existence of native oxides in metal precursors can alter the metal-to-boron ratios, resulting in excessive boron (or insufficient metal) in the system [16]. The tolerance of metal vacancies in the hexaboride structure may prevent the formation of boron-rich secondary phases. The existence of metal vacancies (due to native oxides in precursors) is difficult to quantify in the sintered HEHB systems. However, based on the results of HETB systems [16] where the same precursors and fabrication route were utilized, the amount of metal vacancies in HEHBs (or equivalently, excessive boron resulted in boron-rich phase in HETBs) should not be significant. Moreover, all sintered specimens exhibit dark steel blue hue with iridescent tarnish after polishing, an appearance observed for stoichiometric binary hexaborides (with an exception of red-violet $LaB_6$) [35]. This color also implies insignificant metal vacancies, as severely metal-deficient binary hexaboride samples are shown to be blue-gray to gray [35].

Table 1 also includes the measured densities (via Archimedes method) and the theoretical densities (based on XRD measured lattice parameters and nominal compositions) of all sintered specimens. High relative densities of >98.5% have been demonstrated for all these specimens. Furthermore, SEM micrographs of polished surfaces (Supplementary Fig. S2) with barely 1-2% black spots of porosity and/or unreacted boron also confirm the high relative densities for different specimens.

The homogenous elemental distributions of metal cations in specimen HEHB1 and HEHB3 are confirmed by both microscale SEM-EDS elemental maps and nanoscale STEM-EDS elemental maps (Fig. 3), whereas the elemental homogeneities of other specimens at microscale can be verified by the SEM-EDS elemental maps in Supplementary Fig. S4. AC-STEM high-angle annular dark-field (HAADF) imaging at high magnifications also illustrates the $CaB_6$-typed hexaboride solid solution (for both HEHB1 and HEHB3) at atomic level in Fig. 3(a1) and (b1). In Fig. 3(a1), STEM HAADF image on [001] zone axis demonstrates the cubic structure of the specimen with lattice parameter $a \approx 0.41$ nm, and in Fig. 3(b1), the [211] zone axis and two perpendicular atomic planes $(01\bar{1})$ and $(1\bar{1}\bar{1})$ are marked accordingly. Supplementary Fig. S3 provides the corresponding fast Fourier transform (FFT) diffraction patterns of these two specimens.

SEM-EDS quantitative analyses are utilized to determine the cation compositions of sintered specimens at microscale, and the results are listed in Table 1. Comparing with the nominal equimolar compositions, the measured compositions are different by 1-3%, which are within the typical EDS measurement errors. Hence, the equimolar compositions are confirmed for the sintered specimens and are adopted for the discussion in this study.

Combining all the results above from XRD, SEM, AC-STEM, and EDS, it is clearly demonstrated that a group of $CaB_6$-typed cubic rare earth HEHB solid solutions have been successfully synthesized to fully dense (relative density >98.5%) in bulk pellets via in-situ metal-boron reactive SPS. Since the previously reported rare earth HEHBs (prepared by thermal reduction of rare earth metal oxides) are synthesized in forms of powders [33] or highly porous materials [13], the successful fabrication of HEHBs in this study epitomizes the first group of dense HEHBs in the bulk form. Moreover, previous studies only involve Y, Ce, Nd, Sm, Eu, Er and Yb (for powders) [33] as well as Y, Nd, Sm, Eu, and Yb (for porous materials) [13], this study further successfully incorporates La, Pr, Gd, Tb, and Dy into HEHB systems for the first time.

Performed on polished specimen surfaces normal to the direction of pressure and current during SPS, EBSD analyses are utilized to measure the grain size and examine the texture of all sintered specimens. Normal direction inverse pole figure orientation maps for all HEHB specimens are illustrated in Fig. 4 with insets of corresponding grain size distributions. The average grain sizes (± one standard deviations) of specimen HEHB1 to HEHB6 are measured to be 4.72 ± 32.87 µm, 4.12 ± 2.48 µm, 4.22 ± 2.10 µm, 4.62 ± 2.50 µm, 5.05 ± 2.83 µm, and 3.48 ± 2.56 µm, respectively. All these sintered HEHB specimens exhibit similar averaged grain sizes of ~3.5-5.0 µm due to the same synthesizing route of HEMB succeeded by SPS at 1700 °C applied. Noticeably, specimen HEHB6 contains a small number of large grains and some clusters of tiny grains, which is commonly observed in borides prepared by in-situ metal-boron reactive SPS [16, 38, 43, 44]. All sintered HEHB specimens show no noticeable texture in the grain orientation maps, which can also be verified by the consistency in relative peak intensities between the measured and calculated XRD patterns (Fig. 2 vs. Fig. S1).

Vickers microhardness tests have been performed on polished specimen surfaces at a standard indentation load of 9.8 N; the results are listed in Table 1. All sintered HEHB specimens demonstrate consistent Vickers microhardness of 16-18 GPa, which are harder than the HETBs (13-15 GPa) [16] fabricated via the same route. The increased hardness vales in HEHBs can be attributed to the fact that tetragonal tetraborides are less rigid along [001] than the octahedra boron chains in hexaborides [39]. Nevertheless, these HEHBs are still noticeably softer than high-entropy diborides [45-47] and $M_3B_4$ borides [15], as well as the superhard solid-solution hexagonal tetraborides [48, 49], dodecaborides [50] and high-entropy monoborides [14, 51].

In addition, nanoindentation tests conducted at 500 mN measured the nanoindentation hardness values between 19-22 GPa (Table 1). The nanohardness values are 10-30% higher than those of Vickers microhardness measured at 9.8 N, which is reasonable due to the indentation size effect [52] at different indentation loads as well as the different Oliver and Pharr method [53] adopted for nanoindentation hardness analysis (that utilizes projected contact area at peak load, instead of the residual projected area, and assumes purely elastic contact). For example, nanoindentation hardness has been reported to be ~20% higher than the microhardness (at the same indentation load) for Si and fused silica [54]. The measured hardness values via both routes for all six specimens are also illustrated in Fig. 5(a).

The microhardness of binary hexaborides have been widely investigated (via different methods and loading conditions) and summarized in a handbook [35], as well as a recent review [31]. Among them, Binder [55] systematically investigated the hardness of most rare earth hexaborides on polycrystalline samples via Knoop microhardness test at 0.98 N; and their measured hardness are all within the range of ~18-21 GPa. Considering that Knoop microhardness test generally gives a similar (but slightly lower) measured value than Vickers microhardness at the same indentation loads on ceramic materials [56], the hardness of the HEHBs are roughly comparable to the RoM average of constituent binary hexaborides. It should also be noted that the hardness of some binary hexaborides have also been reported to be 23-27 GPa by Vickers microhardness test at 0.98 N or lower in early reports [31, 35]. On the other hand, limited reports on nanohardness of binary hexaborides have been found. $LaB_6$ has been reported to possess a nanoindentation hardness of ~23 GPa at 50 mN [57], whereas a higher indentation hardness of ~35 GPa has been found at a lower load of 10 mN [58]. These results are not surprising given the difference in nanoindentation loads.

Young's moduli are also determined to be 190-250 GPa by nanoindentation tests for all six specimens (illustrated in Fig. 5(b) and listed in Table 1). Higher values of 376 GPa [57] and 393 GPa [58] have been reported for $LaB_6$ at lower nanoindentation loads of 50 mN and 10 mN, respectively. Besides the difference in nanoindentation loads, this big discrepancy of Young's moduli between the HEHB and $LaB_6$ specimens might also related to the melting process during the fabrication of $LaB_6$ specimens. While some early studies also reported the Young's moduli for binary hexaborides to be 350-400 GPa [35], a recent study in 2018 reported the Young's modulus of hot-pressed $SmB_6$ to be 271 GPa by fracture strength test and 244 GPa by ultrasonic test [59], which are more comparable with our HEHBs.

UPS measurements were performed on freshly polished specimen surfaces, with the resultant spectra shown in Fig. 6. Work functions have been determined based on the incident photon energy of He I emission (21.22 eV) and the respective high-binding-energy cut-off (marked by the black vertical line in Fig. 6) of each specimen. All HEHB specimens demonstrate comparable measured work functions within 3.7-4.0 eV (listed in Table 1), which are noticeably higher than the constituent $LaB_6$ (that commonly used as hot cathode material) work function of ~2.6-2.8 eV on (100) [37, 60]. Various early studies also demonstrated the work functions of other constituent binary hexaborides to be ~2.9-3.5 eV with results summarized in Ref. [35] (except for $YB_6$ that experiences severe vaporization at elevated temperature [61]). It should be pointed out that most of the reported work functions are obtained by thermionic emission and Richardson's law. Nevertheless, the method of photoelectric yield should generate consistent results, as similar work function results have been obtained from both methods for $LaB_6$ [62] (also found in Ref. [35]). In this scenario, it can be deduced that the work functions of HEHBs are higher than the RoM average of their constituent binary hexaborides, which implies that they might not be desirable for hot cathode application.

4. **Conclusions**

A group of $CaB_6$-typed cubic rare earth high-entropy hexaborides or HEHBs have been successfully synthesized into bulk forms via HEBM followed by direct metal-boron reactive SPS. The sintered bulk HEHB specimens are >98.5% in relative densities without detectable oxides or other secondary boride phases. The homogenous distributions of metal cations have been verified by both SEM-EDS at microscale and STEM-EDS at nanoscale. The averaged grain sizes have been determined to be between 3.5-5.0 μm. Vickers microhardness has been measured to be 16-18 GPa at a standard load of 9.8 N. The nanoindentation hardness and Young's moduli have been determined to be 19-22 GPa and 190-250 GPa, respectively, by nanoindentation tests using a maximum load of 500 mN. Noticeably, work functions of these sintered HEHB specimens have been determined to be 3.7-4.0 eV by UPS characterizations, which are significantly higher the work function of $LaB_6$.

In contrast to the previously reported rare earth HEHB in powders [33] or highly porous materials [13], HEHBs have been fabricated in bulk form for the first time in this study. This added a new class to bulk HECs and enabled us to measure some basic bulk properties of HEHBs.

**Acknowledgement:**

We acknowledge the support by the UC Irvine MRSEC, Center for Complex and Active Materials, under National Science Foundation (NSF) Grant No. DMR-2011967 and the earlier support from an ONR MURI program (N00014-15-1-2863). We thank Kim Kisslinger for his help with the TEM experiments. UPS work was performed using instrumentation funded in part by the National Science Foundation Major Research Instrumentation Program under grant CHE-1338173. We thank Ich Tran for his help with the UPS experiment. This research used shared characterization facilities of UCSD NanoEngineering's Materials Research Center (NE-MRC), the San Diego Nanotechnology Infrastructure (SDNI) of UCSD, a member of the National Nanotechnology Coordinated Infrastructure supported by the NSF (ECCS-1542148), UC Irvine Materials Research Institute (IMRI), which is also supported in part by the NSF (DMR-2011967), and the Center for Functional Nanomaterials, which is a U.S. DOE Office of Science Facility, at Brookhaven National Laboratory (DE-SC0012704).

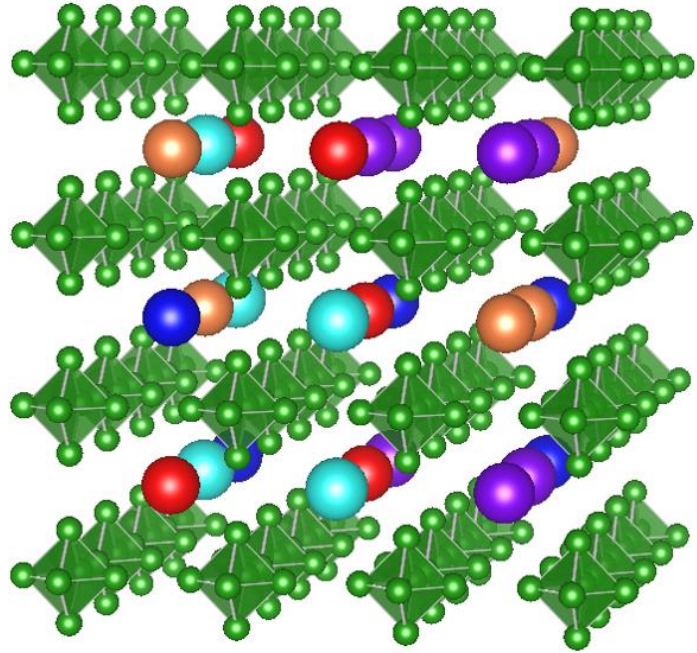

**Fig. 1.** Schematic illustration of atomic structure of the $CaB_6$-prototyped cubic high-entropy hexaboride (HEHB, Space group: Pm-3m, No. 221). In its cubic unit cell, metal cations locate at the body-center and octahedra composed of six boron atoms situate at eight vertices. This $CaB_6$-typed structure can also be viewed as CsCl-typed with $B_6$ octahedra occupying the anion positions. Here, small green balls denote boron atoms, and colored large balls represent atoms of five different rare earth elements selected from Y, La, Pr, Nd, Sm, Gd, Tb, Dy, and Yb.

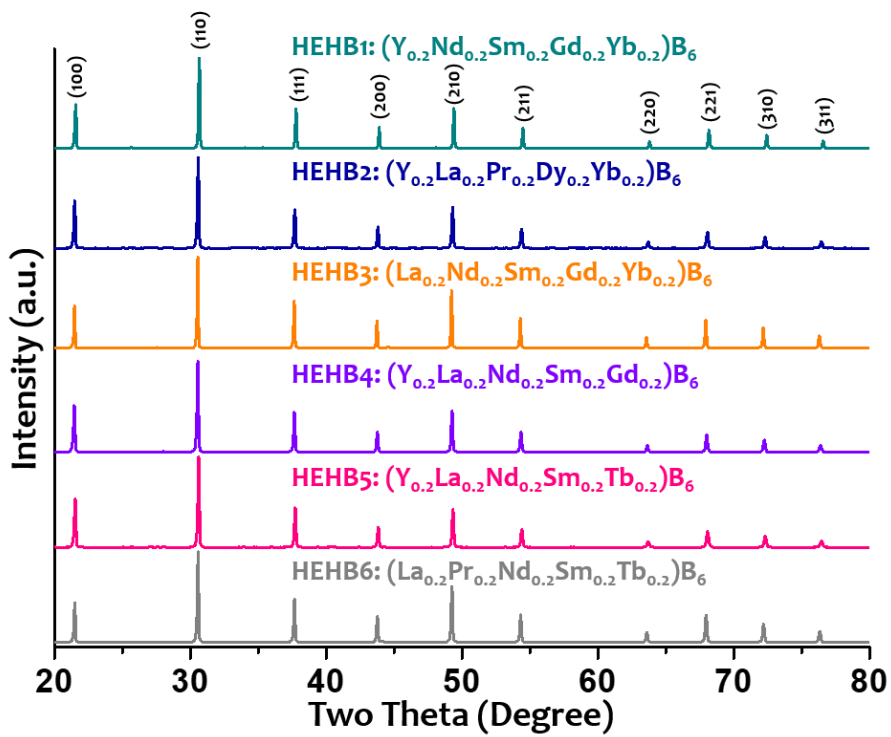

**Fig. 2.** XRD patterns of specimens HEHB1 to HEHB6 fabricated via in-situ (direct) metal-boron reactive SPS. All six specimens synthesized in this study exhibit a single $CaB_6$-prototyped cubic hexaboride phase without any detectable oxides or any other secondary phase.

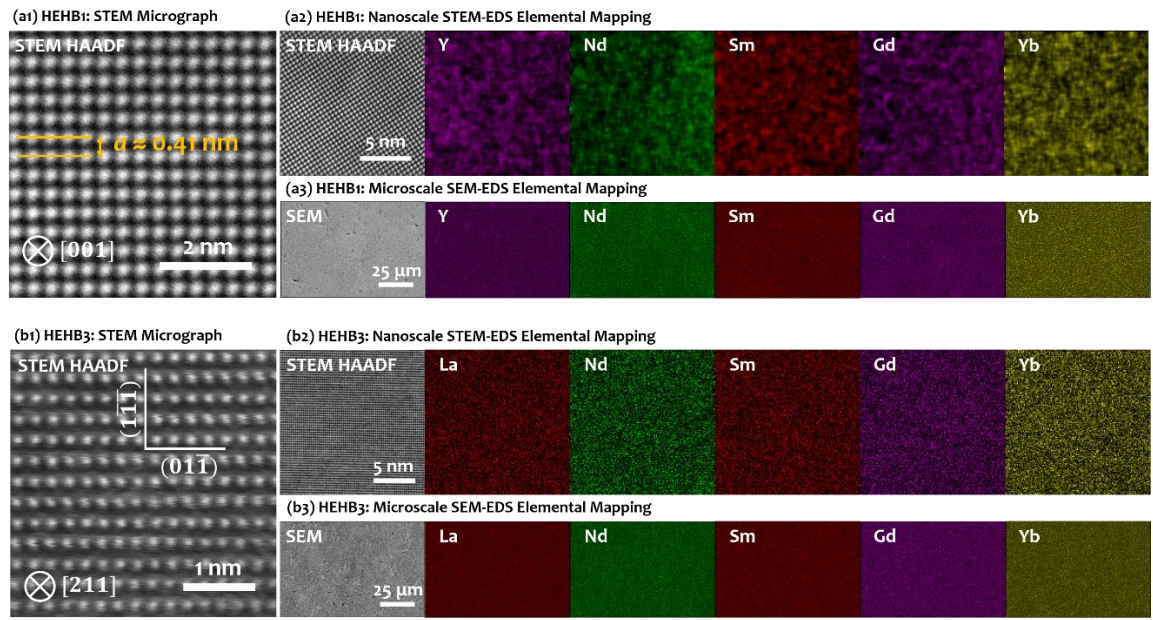

**Fig. 3.** STEM micrographs of specimen **(a1)** HEHB1 $(Y_{0.2}Nd_{0.2}Sm_{0.2}Gd_{0.2}Yb_{0.2})B_6$ and **(b1)** $(La_{0.2}Nd_{0.2}Sm_{0.2}Gd_{0.2}Yb_{0.2})B_6$. The STEM HAADF image at high magnification illustrates the cation atomic structure. For HEHB1, the [001] zone axis and the lattice parameter $a$ ($\approx 0.41$ nm) are labeled. For HEHB3, the [211] zone axis and two perpendicular atomic planes $(01\bar{1})$ and $(1\bar{1}\bar{1})$ are marked. See Supplementary Fig. S3 for the FFT diffraction patterns of these two specimens for further illustration of the crystallographic orientation. STEM micrographs with the corresponding EDS elemental maps of specimen **(a2)** HEHB1 and **(b2)** HEHB3, together with the SEM micrographs with the EDS elemental maps of the same specimens **(a3)** and **(b3)**, demonstrate the homogenous elemental distributions at both nanoscale and microscale. See Supplementary Fig. S4 for their SEM-EDS elemental maps for the other four specimens in this study that have confirmed the compositional homogeneities.

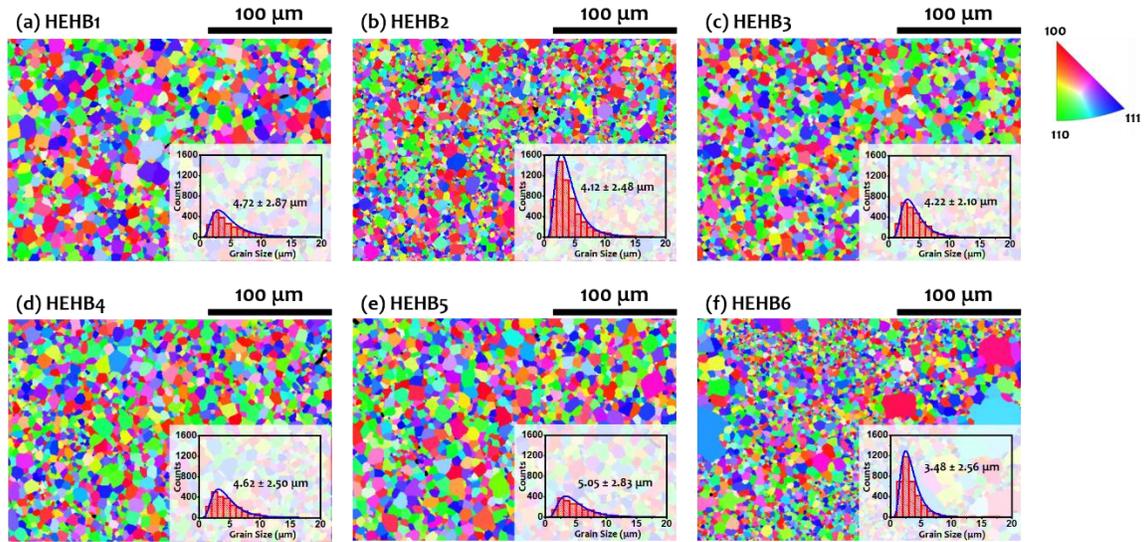

**Fig. 4.** EBSD normal direction inverse pole figure orientation maps for all six synthesized specimens: **(a)** HEHB1, **(b)** HEHB2, **(c)** HEHB3, **(d)** HEHB4, **(e)** HEHB5, and **(f)** HEHB6. All specimens are fabricated via the same route of in-situ metal-boron reactive SPS. All specimens show a similar averaged grain size of ~3-5 μm.

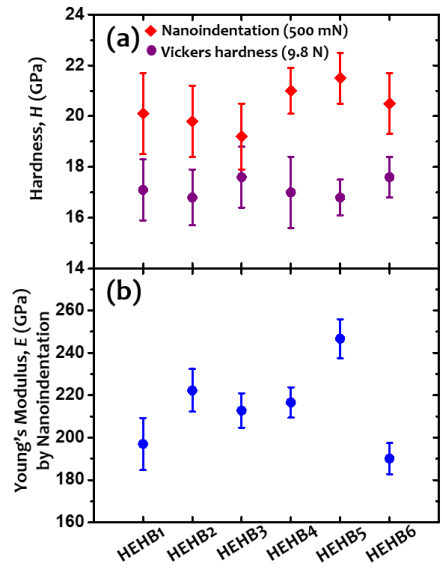

**Fig. 5. (a)** Hardness and **(b)** Young's moduli for the specimens synthesized in this study. The hardness is measured by both nanoindentation (at 500 mN) and Vickers microhardness test (at 9.8 N). The Young's moduli are obtained by nanoindentation.

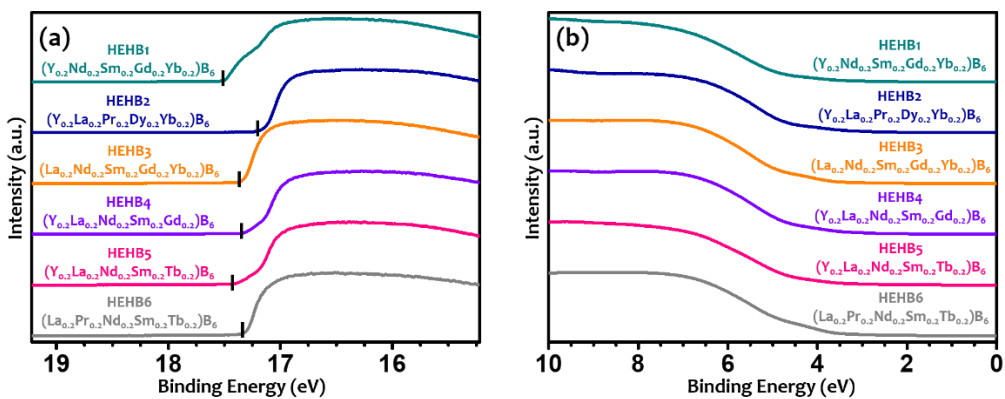

**Fig. 6.** UPS spectra of specimens HEHB1 to HEHB6: **(a)** high-binding-energy cut-off region and **(b)** low-binding energy region. The work function is calculated utilizing the high-binding-energy cut-off, marked by the black vertical line, for each specimen.

**Table 1. Summary of the six specimens (HEHB1 to HEHB6) synthesized in this study.** All specimens were fabricated via the same routine of in-situ metal-boron reactive SPS. The measured compositions for each specimen were directly attained from SEM-EDS analyses. The lattice parameters were measured by XRD spectra. Averaged lattice parameters were calculated via the rule of mixture (RoM) on weighted means of binary hexaborides (individual metal hexaborides). Theoretical densities were calculated from the measured lattice parameters and the nominal compositions for each specimen. The densities were measured experimentally via Archimedes method. Grain size information was obtained from EBSD analyses. Vickers microhardness were measured at indentation load of 9.8 N (1 kgf), whereas the nanoindentation hardness were measured at 500 mN. Young's moduli were also obtained by nanoindentation. Work functions were calculated by the incident photon energy of He I emission ($h\nu$ = 21.22 eV) and the high-binding-energy cut-offs ($E_{cut-off}$) of each specimen via $WF = h\nu - E_{cut-off}$.

| Specimen | Nominal Compositions | Measured Compositions | Measured Lattice Parameter $a$ by XRD (Å) | Averaged Lattice Parameter $a$ via RoM (Å) | Theoretical Density (g/cm³) | Measured Density (g/cm³) | Relative Density | Grain Size (μm) | Vickers Microhardness at 9.8 N (GPa) | Nanoindentation hardness at 500 mN (GPa) | Young's Modulus (GPa) | Work Function (eV) |
|---|---|---|---|---|---|---|---|---|---|---|---|---|
| HEHB1 | $(Y_{0.2}Nd_{0.2}Sm_{0.2}Gd_{0.2}Yb_{0.2})B_6$ | $(Y_{0.23}Nd_{0.21}Sm_{0.20}Gd_{0.18}Yb_{0.18})B_6$ | 4.1223 | 4.1266 | 4.92 | 4.87 | 99.0% | 4.72 ± 2.87 | 17.1 ± 1.2 | 20.1 ± 1.6 | 197.0 ± 12.3 | 3.71 |
| HEHB2 | $(Y_{0.2}La_{0.2}Pr_{0.2}Dy_{0.2}Yb_{0.2})B_6$ | $(Y_{0.19}La_{0.19}Pr_{0.22}Dy_{0.19}Yb_{0.21})B_6$ | 4.1288 | 4.1294 | 4.85 | 4.78 | 98.6% | 4.12 ± 2.48 | 16.8 ± 1.1 | 19.8 ± 1.4 | 222.3 ± 10.0 | 4.00 |
| HEHB3 | $(La_{0.2}Nd_{0.2}Sm_{0.2}Gd_{0.2}Yb_{0.2})B_6$ | $(La_{0.19}Nd_{0.21}Sm_{0.23}Gd_{0.19}Yb_{0.18})B_6$ | 4.1351 | 4.1350 | 5.11 | 5.07 | 99.2% | 4.22 ± 2.10 | 17.6 ± 1.2 | 19.2 ± 1.3 | 212.8 ± 8.1 | 3.84 |
| HEHB4 | $(Y_{0.2}La_{0.2}Nd_{0.2}Sm_{0.2}Gd_{0.2})B_6$ | $(Y_{0.22}La_{0.20}Nd_{0.19}Sm_{0.22}Gd_{0.17})B_6$ | 4.1322 | 4.1280 | 4.73 | 4.69 | 99.2% | 4.62 ± 2.50 | 17.0 ± 1.4 | 21.0 ± 0.9 | 216.6 ± 7.2 | 3.87 |
| HEHB5 | $(Y_{0.2}La_{0.2}Nd_{0.2}Sm_{0.2}Tb_{0.2})B_6$ | $(Y_{0.21}La_{0.17}Nd_{0.18}Sm_{0.21}Tb_{0.23})B_6$ | 4.1283 | 4.1260 | 4.75 | 4.68 | 98.5% | 5.05 ± 2.83 | 16.8 ± 0.7 | 21.5 ± 1.0 | 246.7 ± 9.2 | 3.80 |
| HEHB6 | $(La_{0.2}Pr_{0.2}Nd_{0.2}Sm_{0.2}Tb_{0.2})B_6$ | $(La_{0.18}Pr_{0.21}Nd_{0.19}Sm_{0.21}Tb_{0.21})B_6$ | 4.1341 | 4.1302 | 4.97 | 4.92 | 99.0% | 3.48 ± 2.56 | 17.6 ± 0.8 | 20.5 ± 1.2 | 190.1 ± 7.3 | 3.86 |